\documentclass[10pt]{article}
\usepackage{graphicx}
\usepackage{amsmath}
\usepackage{amssymb}
\usepackage{caption2}
\begin{document}
\begin{center}
{\large {\bf \sc{   Triply-charmed dibaryon states or two-baryon scattering states from the QCD sum rules?
  }}} \\[2mm]
Zhi-Gang  Wang \footnote{E-mail: zgwang@aliyun.com.  }   \\
 Department of Physics, North China Electric Power University, Baoding 071003, P. R. China
\end{center}

\begin{abstract}
In this article,  we construct the color-singlet-color-singlet type currents to study the scalar and axialvector $\Xi_{cc}\Sigma_c$ dibaryon states with
QCD sum rules in details by  taking
 into account both the dibaryon states and two-baryon scattering sates at the hadron side,  and examine the existence of the $\Xi_{cc}\Sigma_c$ dibaryon states.
 Our calculations indicate that the two-baryon scattering states  cannot saturate the QCD sum rules,  it is necessary to introduce the dibaryon states, the color-singlet-color-singlet type currents couple potentially to the molecular states,  not to the two-particle scattering states, the molecular states begin to receive contributions at the order $\mathcal{O}(\alpha_s^0)$, not at the order $\mathcal{O}(\alpha_s^2)$.
\end{abstract}



\section{Introduction}

A dibaryon state denotes an object with baryon number $2$,  the oldest known dibaryon state  is the deuteron, which is  a very loosely
 bound state of two baryons, a proton and a neutron, and is made of six light valence quarks.
In 2014, the  WASA-at-COSY collaboration established the narrow resonance structure $d^*(2380)$ with $I(J^P) = 0(3^+)$ as a genuine $s$-channel
resonance using partial-wave analysis, and given  the first clear-cut experimental evidence
for the existence of a true
dibaryon resonance \cite{d2380-establish}, the $d^*(2380)$ was firstly observed in the double-pionic fusion to the deuteron \cite{d2380-pipi}.
The $d^*(2380)$ may be a $\Delta\Delta$ dibaryon state or a six-quark state, for more theoretical and experimental  works on the  light flavor dibaryon states, one can consult  the comprehensive review \cite{dibaryon-review}.

On the other hand, many  charmonium-like and bottomonium-like states were observed after the discovery of the $X(3872)$ by the  Belle collaboration in 2003 \cite{X3872-2003}. It is difficult to accommodate those exotic $X$, $Y$ and $Z$ states in the conventional meson spectrum, especially, the charged
charmonium-like states are  good candidates for the  multiquark states, tetraquark states or molecular states \cite{GFK-RMP}.

The observation of the $d^*(2380)$  triggers much theoretical interest  on  possible existence of the
molecular states made of two heavy baryons. As the large masses of the heavy baryons reduce the kinetic energy of the two-baryon systems, which makes it easier to form bound states.
In 2015,  the  LHCb collaboration  observed  two pentaquark candidates $P_c(4380)$ and $P_c(4450)$ in the $J/\psi p$ mass spectrum \cite{LHCb-4380}.
In 2019, the LHCb collaboration observed a  narrow pentaquark candidate $P_c(4312)$ and confirmed the $P_c(4450)$ pentaquark structure, which consists  of two narrow overlapping peaks $P_c(4440)$ and $P_c(4457)$ \cite{LHCb-Pc4312}. The $P_c(4380)$, $P_c(4440)$ and $P_c(4457)$  lie near the $\bar{D}\Sigma_c$, $\bar{D}\Sigma_c^*$, $\bar{D}^{*}\Sigma_c$  and   $\bar{D}^{*}\Sigma_c^*$ thresholds respectively, which leads to the conjecture that they are meson-baryon type molecular states.
We can  learn something  about the $\bar{D}\Sigma_c$, $\bar{D}\Sigma_c^*$, $\bar{D}^{*}\Sigma_c$  and   $\bar{D}^{*}\Sigma_c^*$ pentaquark molecular states from
the $\Xi_{cc} \Sigma_c$, $\Xi_{cc} \Sigma_c^*$, $\Xi_{cc}^* \Sigma_c$ and $\Xi_{cc}^* \Sigma_c^*$ dibaryon states (or vice versa), which
  are connected with each other by heavy antiquark-diquark symmetry, if we assume the light-quark structures are almost
  identical \cite{PanLDXX}. In Ref.\cite{ccc-dibaryon}, the molecular states consist of a
doubly charmed baryon  and an S-wave charmed baryon are investigated in the one-boson-exchange model.

Exploring the hadron-hadron interactions  plays  an important role in understanding the meson-meson type, meson-baryon type, baryon-antibaryon type, baryon-baryon type molecular states. It is essential to make a detailed theoretical investigation of those molecular states to encourage or stimulate  new experiments to search for some  evidences. Theoretically, we can study the molecular states at the hadron level \cite{PanLDXX,ccc-dibaryon,Heavy-dibaryon-hadron}
or at the quark level \cite{Latt-dibaryon-PRL,Heavy-dibaryon-quark}.

In Ref.\cite{Latt-dibaryon-PRL}, Junnarkar and  Mathur study the  $\Sigma_c\Xi_{cc}$, $\Omega_c\Omega_{cc}$,
$\Sigma_b\Xi_{bb} $, $\Omega_b\Omega_{bb} $ and $\Omega_{ccb}\Omega_{cbb}$ dibaryon states with  $J^P=1^+$ via the lattice QCD,
and unambiguously find that the ground state masses of the dibaryon states $\Omega_c\Omega_{cc}$, $\Omega_b\Omega_{bb} $ and
$\Omega_{ccb}\Omega_{cbb} $ are below their respective
two-baryon thresholds, but cannot obtain definitive conclusion   about the existence  of the
$\Sigma_c\Xi_{cc}$ and $\Sigma_b\Xi_{bb}$ dibaryon states due to large systematic errors. The doubly or triply heavy baryon states have not been observed experimentally  yet except for the $\Xi_{cc}$ \cite{LHCb-Xicc}.  So it is interesting to study the $\Sigma_c\Xi_{cc}$ dibaryon states with the QCD sum rules, as we have experimental input  to assess the bound energies.
In Ref.\cite{Oka-H-dibaryon}, the $H$-dibaryon or $\Lambda\Lambda$ dibaryon state is studied with the QCD sum rules.
In Ref.\cite{D2380-CHHX}, the $d^*(2380)$ is assigned as a $\Delta\Delta$ dibaryon state and studied with the QCD sum rules.

The QCD sum rules approach is a powerful nonperturbative  tool in studying the ground state tetraquark states, tetraquark molecular states,  pentaquark states, pentaquark molecular states, and has given many successful descriptions of  the hadronic parameters, such as the masses and decay widths \cite{QCDSR-4-quark-mass,WangHuangtao-PRD,QCDSR-4-quark-width,WangZG-4-quark-mole,WangZG-CPC-Y4390,QCDSR-5-quark-mole,QCDSR-WangZG-5-quark-mole,QCDSR-5-quark-penta}.
However, different voice arises, Lucha, Melikhov and Sazdjian assert   that the QCD sum rules are  misused to study the tetraquark or molecule masses,  all the four-quark currents can be rearranged into the color-singlet-color-singlet type currents, the contributions  at the order $\mathcal{O}(\alpha_s^k)$ with $k\leq1$ in the operator product expansion, which are factorizable in the color space, are exactly  canceled out   by the two-meson scattering states at the hadron side, the tetraquark molecular states begin to receive contributions at the order $\mathcal{O}(\alpha_s^2)$ \cite{Chu-Sheng-PRD,Chu-Sheng-EPJC}. The tetraquark molecular states are two-meson bound states or resonant states in contrast to the two-meson scattering states.
 The dibaryon states consist of two color-singlet objects, they are an another type molecular states. In this article, we will examine the applicability  of the QCD sum rules to study the two-baryon type molecular states.

 In this article, we study the scalar and axialvector $\Xi_{cc}\Sigma_c$ dibaryon states with QCD sum rules in details. We take into account both the
  dibaryon states and two-baryon scattering sates at the hadron side, and examine whether or not the QCD sum rules support the existences of dibaryon states.

The article is arranged as follows:  we obtain the QCD sum rules for the masses and pole residues of  the
 triply-charmed dibaryon states and examine the applicability  of the QCD sum rules in Sect.2;  in Sect.3, we present the numerical results and discussions; and Sect.4 is reserved for our
conclusion.

\section{QCD sum rules for  the dibaryon states}

In the following, we write down  the two-point correlation functions $\Pi_S(p)$ and $\Pi_{\mu\nu}(p)$ in the QCD sum rules,
\begin{eqnarray}
\Pi_S(p)&=&i\int d^4x e^{ip \cdot x} \langle0|T\Big\{J(x)J^{\dagger}(0)\Big\}|0\rangle \, , \nonumber\\
\Pi_{\mu\nu}(p)&=&i\int d^4x e^{ip \cdot x} \langle0|T\Big\{J_{\mu}(x)J^{\dagger}_{\nu}(0)\Big\}|0\rangle \, ,
\end{eqnarray}
where
 \begin{eqnarray}
 J(x)&=&  J^T_c(x) C\gamma_5 J_{cc}(x) \, , \nonumber\\
 J_\mu(x)&=&  J^T_c(x) C\gamma_\mu J_{cc}(x) \, , \nonumber\\
 J_{c}(x)&=& \varepsilon^{ijk} q^T_i(x) C\gamma_\alpha q_{j}(x)\gamma^\alpha\gamma_5c_k(x) \, , \nonumber\\
  J_{cc}(x)&=& \varepsilon^{ijk} c^T_i(x) C\gamma_\alpha c_{j}(x)\gamma^\alpha\gamma_5q_k(x) \, ,
\end{eqnarray}
 the $i$, $j$ and $k$ are color indexes, the $C$ is the charge conjugation matrix. The currents $J(x)$ and $J_\mu(x)$ have two color-neutral clusters and have the property under
 parity transformation,
 \begin{eqnarray}
\widehat{P}J(x)\widehat{P}^{-1}&=&+ J(\tilde{x}) \, , \nonumber\\
\widehat{P}J_{\mu}(x)\widehat{P}^{-1}&=&- J^\mu(\tilde{x}) \, ,
\end{eqnarray}
where the coordinates $x^\mu=(t,\vec{x})$ and $\tilde{x}^\mu=(t,-\vec{x})$.

We construct the Ioffe's currents $J_c(x)$ and $J_{cc}(x)$ according to the Fermi-Dirac statistics and the attractive interactions originate from the one-gluon exchange.
The currents $J_c(x)$ and $J_{cc}(x)$  have the $J^P={\frac{1}{2}}^+$ and have the constituent quarks or valence  quarks  $qqc$ and $ccq$, respectively, just like the baryon states
$\Sigma_c$ and $\Xi_{cc}$. The quantum field theory does not forbid the current-baryon couplings,
\begin{eqnarray}
\langle 0|J_c(0)|\Sigma_c(p)\rangle&=&\lambda_{\Sigma}\,U_{\Sigma}(p)\, , \nonumber\\
\langle 0|J_{cc}(0)|\Xi_{cc}(p)\rangle&=&\lambda_{\Xi}\,U_{\Xi}(p)\, ,
\end{eqnarray}
the $U_\Sigma(p)$ and $U_{\Xi}(p)$ are Dirac spinors.
The values of the coupling constants or pole residues $\lambda_{\Sigma}$ and
$\lambda_{\Xi}$ are not experimentally  measurable quantities, we have to calculate those values with the QCD sum rules or lattice QCD.

The $\Sigma_c\Xi_{cc}$ systems maybe form the $\Sigma_{c}\Xi_{cc}$ dibaryon states, or maybe not form the $\Sigma_{c}\Xi_{cc}$ dibaryon states (in other words,  they are just the $\Sigma_{c}\Xi_{cc}$ two-baryon scattering states). If they form the $\Sigma_{c}\Xi_{cc}$ dibaryon states,
the quantum field theory does not forbid the current-dibaryon couplings, the currents $J(x)$ and $J_\mu(x)$ couple potentially to the
scalar and axialvector dibaryon states,  respectively, or to the two-baryon scattering states with the spin-parity $J^P=0^+$ and $1^+$, respectively,
\begin{eqnarray}
\langle 0|J(0)|D_S(p)\rangle&=&\lambda_{S}\, , \nonumber\\
\langle 0|J_\mu(0)|D_A(p)\rangle&=&\lambda_{A}\varepsilon_\mu\, , \nonumber\\
\langle 0|J(0)|\Sigma_c(q)\Xi_{cc}(p-q)\rangle&=&\lambda_{\Sigma}\lambda_{\Xi}\,U_\Sigma^T(p)C\gamma_5U_{\Xi}(p-q)\, , \nonumber\\
\langle 0|J_\mu(0)|\Sigma_c(q)\Xi_{cc}(p-q)\rangle&=&\lambda_{\Sigma}\lambda_{\Xi}\,U_\Sigma^T(p)C\gamma_{\mu}U_{\Xi}(p-q)\, ,
\end{eqnarray}
we use the $D$ to denote the dibaryon states.

At the hadron side of the correlation functions $\Pi_S(p)$ and $\Pi_{\mu\nu}(p)$,   we  isolate  the contributions of  both the lowest dibaryon states and  two-baryon scattering states,
\begin{eqnarray}
\Pi_{\mu\nu}(p)&=&\Pi_A(p)\left( -g_{\mu\nu}+\frac{p_{\mu}p_{\nu}}{p^2}\right)+\cdots\, ,
\end{eqnarray}
\begin{eqnarray}
  \Pi_S(p) & = &\frac{ \lambda_{S}^2}{M_{S}^2-p^2}    +\int_{\Delta^2}^{s_0}ds\frac{1}{s-p^2}\rho_{H,S}(s)+\cdots \, , \nonumber\\
   \Pi_{A}(p) & = &\frac{ \lambda_{A}^2}{M_{A}^2-p^2}    +\int_{\Delta^2}^{s_0}ds\frac{1}{s-p^2}\rho_{H,A}(s)+\cdots \, ,
\end{eqnarray}
where
\begin{eqnarray}
\rho_{H,S}&=&\frac{\lambda_\Sigma^2\lambda_\Xi^2}{8\pi^2}\frac{\sqrt{\lambda(s,m_\Sigma^2,m_\Xi^2)}}{s}\left[s-(m_\Xi-m_\Sigma)^2\right]\, , \nonumber\\
\rho_{H,A}&=&\frac{\lambda_\Sigma^2\lambda_\Xi^2}{8\pi^2}\frac{\sqrt{\lambda(s,m_\Sigma^2,m_\Xi^2)}}{s}
\left[s-(m_\Xi-m_\Sigma)^2-\frac{\lambda(s,m_\Sigma^2,m_\Xi^2)}{3s}\right]\, ,
\end{eqnarray}
where $\Delta^2=(m_\Sigma+m_\Xi)^2$, $\lambda(a,b,c)=a^2+b^2+c^2-2ab-2bc-2ac$, the $s_0$ are the continuum threshold parameters.

In the QCD sum rules, we carry out the operator product expansion at the deep Euclidean region $P^2=-p^2\to\infty$, which corresponds to the small spatial distance and
time interval $x_0\sim \vec{x}\sim \frac{1}{\sqrt{P^2}}$, and $x^2\sim \frac{1}{P^2}$, it is questionable to apply the Landau equation to study the Feynman diagrams \cite{Landau}.
At the QCD side of the correlation functions $\Pi_S(p)$ and $\Pi_{\mu\nu}(p)$, we contract the $q$ and $c$ quark fields  with Wick theorem and obtain the results,
\begin{eqnarray}\label{PI-S}
\Pi_S(p)&=&-i\,\varepsilon^{ijk}\varepsilon^{lmn}\varepsilon^{i^{\prime}j^{\prime}k^{\prime}}
\varepsilon^{l^{\prime}m^{\prime}n^{\prime}}\int d^4x\, e^{ip\cdot x} \nonumber\\
&&\left\{  4{\rm Tr}\left[\gamma_\alpha  S^{jj^\prime}(x)   \gamma_{\alpha^\prime} C S^{ii^\prime T}(x)C\right] \,{\rm Tr}\left[\gamma_\beta  S_c^{mm^\prime}(x)   \gamma_{\beta^\prime} C S_c^{ll^\prime T}(x)C\right]  \right.  \nonumber\\
&&\left. {\rm Tr} \left[\gamma_5\gamma^\alpha\gamma_5\gamma^\beta\gamma_5 S^{nn^\prime}(x) \gamma_5 \gamma^{\beta^\prime}\gamma_5\gamma^{\alpha^\prime}\gamma_5 C S_c^{kk^\prime T}(x)C \right]     \right. \nonumber\\
&&-8{\rm Tr}\left[\gamma_\alpha  S^{jj^\prime}(x)   \gamma_{\alpha^\prime} C S^{ii^\prime T}(x)C\right]   \nonumber\\
&& {\rm Tr} \left[\gamma_5\gamma^\alpha\gamma_5\gamma^\beta\gamma_5 S^{nn^\prime}(x) \gamma_5 \gamma^{\beta^\prime}\gamma_5\gamma^{\alpha^\prime}\gamma_5 C S_c^{mk^\prime T}(x)C \gamma_\beta  S_c^{ll^\prime}(x)   \gamma_{\beta^\prime} C S_c^{km^\prime T}(x)C\right]   \nonumber\\
&&-8{\rm Tr} \left[\gamma_5\gamma^\alpha\gamma_5\gamma^\beta\gamma_5 S^{nj^\prime}(x)  \gamma_{\alpha^\prime} C S^{ii^\prime T}(x)C \gamma_\alpha  S^{jn^\prime}(x)\gamma_5   \gamma^{\beta^\prime}\gamma_5\gamma^{\alpha^\prime} \gamma_5 C S_c^{kk^\prime T}(x)C\right] \nonumber\\
&&\, {\rm Tr}\left[\gamma_\beta  S_c^{mm^\prime}(x)   \gamma_{\beta^\prime} C S_c^{ll^\prime T}(x)C\right]   \nonumber\\
&&+16{\rm Tr} \left[\gamma_5\gamma^\alpha\gamma_5\gamma^\beta\gamma_5 S^{nj^\prime}(x)  \gamma_{\alpha^\prime} C S^{ii^\prime T}(x)C \gamma_\alpha  S^{jn^\prime}(x)\gamma_5   \gamma^{\beta^\prime}\gamma_5\gamma^{\alpha^\prime} \gamma_5 C S_c^{mk^\prime T}(x)C\right. \nonumber\\
&&\left.\left.\gamma_\beta  S_c^{ll^\prime}(x)   \gamma_{\beta^\prime} C S_c^{km^\prime T}(x)C\right]\right\} \, ,
\end{eqnarray}

\begin{eqnarray}\label{PI-A}
\Pi_{\mu\nu}(p)&=&i\,\varepsilon^{ijk}\varepsilon^{lmn}\varepsilon^{i^{\prime}j^{\prime}k^{\prime}}
\varepsilon^{l^{\prime}m^{\prime}n^{\prime}}\int d^4x\, e^{ip\cdot x} \nonumber\\
&&\left\{  4{\rm Tr}\left[\gamma_\alpha  S^{jj^\prime}(x)   \gamma_{\alpha^\prime} C S^{ii^\prime T}(x)C\right] \,{\rm Tr}\left[\gamma_\beta  S_c^{mm^\prime}(x)   \gamma_{\beta^\prime} C S_c^{ll^\prime T}(x)C\right]  \right.  \nonumber\\
&&\left. {\rm Tr} \left[\gamma_5\gamma^\alpha\gamma_\mu\gamma^\beta\gamma_5 S^{nn^\prime}(x) \gamma_5 \gamma^{\beta^\prime}\gamma_\nu\gamma^{\alpha^\prime}\gamma_5 C S_c^{kk^\prime T}(x)C \right]     \right. \nonumber\\
&&-8{\rm Tr}\left[\gamma_\alpha  S^{jj^\prime}(x)   \gamma_{\alpha^\prime} C S^{ii^\prime T}(x)C\right]   \nonumber\\
&& {\rm Tr} \left[\gamma_5\gamma^\alpha\gamma_\mu\gamma^\beta\gamma_5 S^{nn^\prime}(x) \gamma_5 \gamma^{\beta^\prime}\gamma_\nu\gamma^{\alpha^\prime}\gamma_5 C S_c^{mk^\prime T}(x)C \gamma_\beta  S_c^{ll^\prime}(x)   \gamma_{\beta^\prime} C S_c^{km^\prime T}(x)C\right]   \nonumber\\
&&-8{\rm Tr} \left[\gamma_5\gamma^\alpha\gamma_\mu\gamma^\beta\gamma_5 S^{nj^\prime}(x)  \gamma_{\alpha^\prime} C S^{ii^\prime T}(x)C \gamma_\alpha  S^{jn^\prime}(x)\gamma_5   \gamma^{\beta^\prime}\gamma_\nu\gamma^{\alpha^\prime} \gamma_5 C S_c^{kk^\prime T}(x)C\right] \nonumber\\
&&\, {\rm Tr}\left[\gamma_\beta  S_c^{mm^\prime}(x)   \gamma_{\beta^\prime} C S_c^{ll^\prime T}(x)C\right]   \nonumber\\
&&+16{\rm Tr} \left[\gamma_5\gamma^\alpha\gamma_\mu\gamma^\beta\gamma_5 S^{nj^\prime}(x)  \gamma_{\alpha^\prime} C S^{ii^\prime T}(x)C \gamma_\alpha  S^{jn^\prime}(x)\gamma_5   \gamma^{\beta^\prime}\gamma_\nu\gamma^{\alpha^\prime} \gamma_5 C S_c^{mk^\prime T}(x)C\right. \nonumber\\
&&\left.\left.\gamma_\beta  S_c^{ll^\prime}(x)   \gamma_{\beta^\prime} C S_c^{km^\prime T}(x)C\right] \right\}\, ,
\end{eqnarray}
where
\begin{eqnarray}\label{LQuarkProg}
S_{ij}(x)&=& \frac{i\delta_{ij}\!\not\!{x}}{ 2\pi^2x^4}-\frac{\delta_{ij}\langle
\bar{q}q\rangle}{12} -\frac{\delta_{ij}x^2\langle \bar{q}g_s\sigma Gq\rangle}{192} -\frac{ig_sG^{a}_{\alpha\beta}t^a_{ij}(\!\not\!{x}
\sigma^{\alpha\beta}+\sigma^{\alpha\beta} \!\not\!{x})}{32\pi^2x^2}  \nonumber\\
&& -\frac{\delta_{ij}x^4\langle \bar{q}q \rangle\langle g_s^2 GG\rangle}{27648} -\frac{1}{8}\langle\bar{q}_j\sigma^{\mu\nu}q_i \rangle \sigma_{\mu\nu}+\cdots \, ,
\end{eqnarray}
\begin{eqnarray}\label{HQuarkProg}
S_c^{ij}(x)&=&\frac{i}{(2\pi)^4}\int d^4k e^{-ik \cdot x} \left\{
\frac{\delta_{ij}}{\!\not\!{k}-m_c}
-\frac{g_sG^n_{\alpha\beta}t^n_{ij}}{4}\frac{\sigma^{\alpha\beta}(\!\not\!{k}+m_c)+(\!\not\!{k}+m_c)
\sigma^{\alpha\beta}}{(k^2-m_c^2)^2}\right.\nonumber\\
&&\left. -\frac{g_s^2 (t^at^b)_{ij} G^a_{\alpha\beta}G^b_{\mu\nu}(f^{\alpha\beta\mu\nu}+f^{\alpha\mu\beta\nu}+f^{\alpha\mu\nu\beta}) }{4(k^2-m_c^2)^5}+\cdots\right\} \, ,\nonumber\\
f^{\alpha\beta\mu\nu}&=&(\!\not\!{k}+m_c)\gamma^\alpha(\!\not\!{k}+m_c)\gamma^\beta(\!\not\!{k}+m_c)\gamma^\mu(\!\not\!{k}+m_c)\gamma^\nu(\!\not\!{k}+m_c)\, ,
\end{eqnarray}
and  $t^n=\frac{\lambda^n}{2}$, the $\lambda^n$ is the Gell-Mann matrix
\cite{WangHuangtao-PRD,Reinders85,Pascual-1984}.
In the full light quark propagator, see Eq.\eqref{LQuarkProg}, we add the term $\langle\bar{q}_j\sigma_{\mu\nu}q_i \rangle$  obtained from  Fierz rearrangement  of the $\langle q_i \bar{q}_j\rangle$ to  absorb the gluons  emitted from other light quark lines and heavy quark lines to extract  the mixed condensate   $\langle\bar{q}g_s\sigma G q\rangle$ \cite{WangHuangtao-PRD}.

\begin{figure}
 \centering
  \includegraphics[totalheight=4cm,width=10cm]{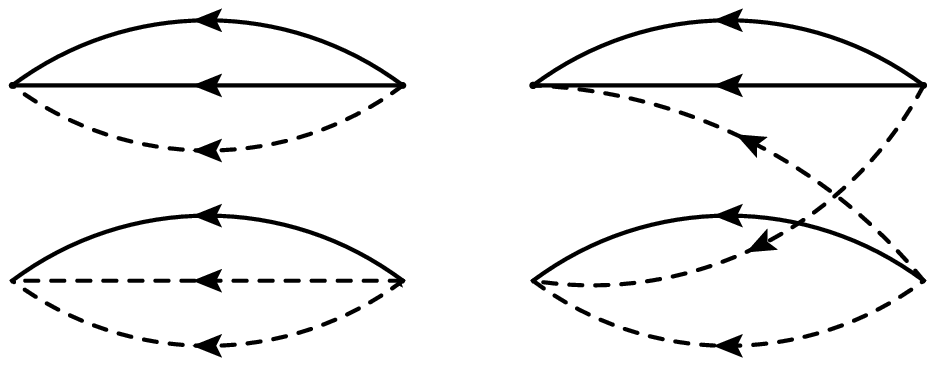}
   \vglue+9mm
   \includegraphics[totalheight=4cm,width=10cm]{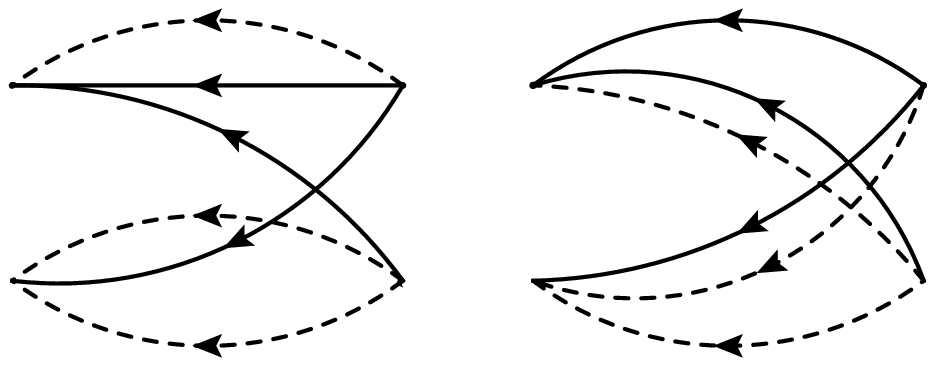}
 \caption{ The  Feynman diagrams  for the lowest order  contributions, where the solid lines and dashed lines represent  the light quarks and heavy quarks, respectively. }\label{Lowest-diagram}
\end{figure}

In Fig.\ref{Lowest-diagram}, we draw the lowest order Feynman diagrams, which correspond to the perturbative contributions in Eqs.\eqref{PI-S}-\eqref{PI-A}, in other words,  the free-quark contributions in the full quark propagators in  Eqs.\eqref{LQuarkProg}-\eqref{HQuarkProg}.
The first diagram in Fig.\ref{Lowest-diagram} is factorizable or disconnected in the color space, the other diagrams are nonfactorizable or connected in the color space.
The (non)factorizable properties in the color space are independent of the (non)factorizable properties in the momentum space. In fact,  all the Feynman diagrams are nonfactorizable
in the momentum space. Thereafter, we will neglect the  phrase "in the color space" for simplicity when discuss the (non)factorizable properties in most cases.

\begin{figure}
 \centering
  \includegraphics[totalheight=4cm,width=10cm]{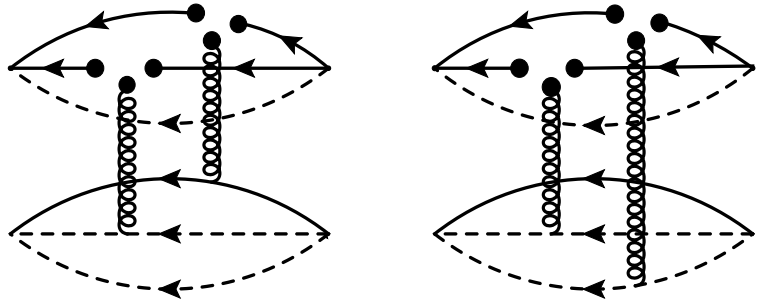}
    \caption{ The  connected Feynman diagrams originate from the first diagram in Fig.\ref{Lowest-diagram}, other
diagrams obtained by interchanging of the heavy quark lines (dashed lines) or light quark lines (solid lines) are implied. }\label{qqg-qqg}
\end{figure}

If we  substitute the light-quark lines and heavy-quark lines in  Fig.\ref{Lowest-diagram} with other terms in the full light-quark and heavy-quark propagators in Eqs.\eqref{LQuarkProg}-\eqref{HQuarkProg}, we can obtain all the relevant Feynman diagrams.
From the first diagram in  Fig.\ref{Lowest-diagram}, we can obtain both connected  and disconnected  Feynman diagrams,  the connected contributions appear due to the   quark-gluon operators  $\bar{q}g_sGq\bar{q}g_sGq$, which are of the order $\mathcal{O}(\alpha_s^1)$ and come from the Feynman diagrams shown in Fig.\ref{qqg-qqg}. From the quark-gluon operators  $\bar{q}g_sGq\bar{q}g_sGq$, we can obtain the vacuum condensate $\langle\bar{q}g_s\sigma Gq\rangle^2$,  the $g_s^2=4\pi \alpha_s$ is absorbed into the vacuum condensate, so the  diagrams in  Fig.\ref{qqg-qqg} can be counted as of the order $\mathcal{O}(\alpha_s^0)$. Those contributions should be taken into account as the QCD sum rules is a nonperturbative method, the connected (or nonfactorizable) Feynman diagrams appear at the order  $\mathcal{O}(\alpha_s^0)$
or $\mathcal{O}(\alpha_s^1)$, not at the order $\mathcal{O}(\alpha_s^2)$ asserted by Lucha, Melikhov and Sazdjian in Refs.\cite{Chu-Sheng-PRD,Chu-Sheng-EPJC}.
 While from other diagrams in  Fig.\ref{Lowest-diagram},  we can obtain only connected (or nonfactorizable) Feynman diagrams in the color space, which appear at the order $\mathcal{O}(\alpha_s^0)$.

 We can use the $\widehat{\Sigma}_c$ and $\widehat{\Xi}_{cc}$ to represent the color-singlet or color-neutral clusters $cqq$ and $ccq$ respectively in the currents.   From those Feynman diagrams, we can observe that the initial color-neutral clusters $\widehat{\Sigma}_c$ and $\widehat{\Xi}_{cc}$ evolve  to
the final color-neutral clusters $\widehat{\Sigma}_c$ and $\widehat{\Xi}_{cc}$ with (without) interchanging colored objects, such as quarks and gluons, in the nonfactorizable (factorizable) Feynman diagrams. The $\Sigma_c$ ($\Xi_{cc}$) and $\widehat{\Sigma}_c$ ($\widehat{\Xi}_{cc}$) have the same quantum numbers $J^P={\frac{1}{2}}^+$, the
quantum field theory allows  nonvanishing couplings between the $\Sigma_c$ ($\Xi_{cc}$) and $\widehat{\Sigma}_c$ ($\widehat{\Xi}_{cc}$), irrespective of   whether  the $\Sigma_c$ and $\Xi_{cc}$ are the two-baryon scattering states or the $\Sigma_c \Xi_{cc}$ components in the dibaryon states.
We cannot obtain other information about the hadrons beyond that from the Feynman diagrams directly,  because we carry out the operator product expansion in  the quantum field theory, there are both perturbative and nonperturbative contributions,  the vacuum condensates are highly nonperturbative quantities and embody the net collective  effects.

 In the similar systems,  the four-quark systems, Lucha, Melikhov and Sazdjian assert  that the contributions  at the order $\mathcal{O}(\alpha_s^k)$ with $k\leq1$ in the operator product expansion, which are factorizable (or disconnected) in the color space, are exactly  canceled out  by the two-meson scattering states at the hadron side, the connected (or nonfactorizable)  Feynman diagrams appear at the order $\mathcal{O}(\alpha_s^2)$, if have  a Landau singularity, begin  to make contributions to the tetraquark state or tetraquark molecular state \cite{Chu-Sheng-PRD,Chu-Sheng-EPJC}.

Without asserting that the factorizable Feynman diagrams in the color space only make contributions to the  two-particle scattering states, the Landau equation
is irrelevant in selecting the Feynman diagrams, because the factorizable Feynman diagrams in the color space can also have  Landau singularities.
  In fact, the quarks and gluons are confined objects, they cannot be put on the mass-shell, it is questionable  to say that the Landau equation is applicably in the nonperturbative  QCD calculations involving bound states \cite{Landau}. Furthermore, even the Landau singularities  appear at the order $\mathcal{O}(\alpha_s^2)$ according to the assertion of Lucha, Melikhov and Sazdjian, we cannot obtain the conclusion that
 those Feynman diagrams make contributions to a tetraquark state or tetraquark molecular state, as the Landau singularity
   only indicates that there exists an intermediate state consists of  four valence  quarks,  irrespective of the tetraquark (molecular) state or two-meson scattering state. Accordingly, in the present case, the Landau singularity only indicates that there exists an intermediate state consists  of  six valence  quarks, irrespective of   the $\Sigma_c\Xi_{cc}$ two-baryon scattering state or the $\Sigma_c \Xi_{cc}$ component in the dibaryon state. The Landau singularity  is just a  kinematical singularity, not a dynamical singularity \cite{GFK-RMP,GuoFK-Hadron2017}.

If we insist on applying  the Landau equation to study the tetraquark molecular states \cite{Chu-Sheng-PRD,Chu-Sheng-EPJC}, we should choose the pole mass to warrant  that there  exists a pole which corresponds to the mass-shell in pure perturbative calculations. In the case of the $c$-quark, the pole mass $\hat{m}_c=1.67\pm0.07\,\rm{GeV}$ from the Particle Data Group \cite{PDG}, $2\hat{m}_c=3.34\pm0.14\,{\rm{GeV}}>m_{\eta_c}$ and $m_{J/\psi}$. It is odd that the charmonium masses lie below the threshold $2\hat{m}_c$ in the QCD sum rules for the $\eta_c$ and $J/\psi$.

 The nonperturbative contributions play an important role in the QCD sum rules, investigating the perturbative contributions of the order $\mathcal{O}(\alpha_s^2)$ alone  cannot lead to feasible QCD sum rules for  the multiquark states.  Furthermore, no feasible QCD sum rules with predictions can be confronted  to the experimental data are obtained according to  the assertion in Refs.\cite{Chu-Sheng-PRD,Chu-Sheng-EPJC}.
 We should bear in mind that the Feynman diagrams at the quark-gluon level involving nonperturbative contributions in the quantum field theory differ greatly from the Feynman diagrams in the  intuitive potential quark models. If we insist on understanding the Feynman diagrams
intuitively, then the disconnected (or factorizable) Feynman diagrams give masses to the $\Sigma_c$ and $\Xi_{cc}$ baryons, the masses are not necessary the physical masses, while the connected (or nonfactorizable) Feynman diagrams contribute attractive interactions to bind the massive $\Sigma_c$ and $\Xi_{cc}$ baryons  to form molecular states.

In Ref.\cite{WangZG-Landau}, I refute the assertion of Lucha, Melikhov and Sazdjian in order and in details, and approve that the Landau equation is of no use in studying the Feynman diagrams in the QCD sum rules for the tetraquark (molecular) states. The assertion is unreasonable, the tetraquark molecular states begin to receive contributions at the order $\mathcal{O}(\alpha_s^0)$ rather than at the order $\mathcal{O}(\alpha_s^2)$.

After computing those Feynman diagrams, we obtain the QCD spectral densities through dispersion relation.
  In this article, we carry out the
operator product expansion to the vacuum condensates  up to dimension-15, and take into account
the  vacuum condensates which are vacuum expectations of the  quark-gluon operators of the order $\mathcal{O}(\alpha_s^k)$ with $k\leq1$.

There are three light quark propagators and three heavy quark propagators in the correlation functions \eqref{PI-S}-\eqref{PI-A}, if each heavy quark line
emits a gluon and each light quark line contributes quark-antiquark pair, we obtain a quark-gluon  operator $g_sG_{\mu\nu}g_sG_{\alpha\beta}g_sG_{\lambda\tau}\bar{q}q\bar{q}q\bar{q}q$, which is of dimension 15, and can lead to the vacuum condensates
$\langle\frac{\alpha_sGG}{\pi}\rangle\langle\bar{q}q\rangle^2\langle\bar{q}g_s\sigma Gq\rangle$, $\langle g_s^3 GGG\rangle\langle\bar{q}q\rangle^3$ and $\langle\bar{q}g_s\sigma Gq\rangle^3$, we take into account the vacuum condensate $\langle\bar{q}g_s\sigma Gq\rangle^3$ and neglect the vacuum condensates $\langle g_s^3 GGG\rangle\langle\bar{q}q\rangle^3$ and $\langle\frac{\alpha_sGG}{\pi}\rangle\langle\bar{q}q\rangle^2\langle\bar{q}g_s\sigma Gq\rangle$. Compared to the $\langle\bar{q}g_s\sigma Gq\rangle^3$, the $\langle g_s^3 GGG\rangle\langle\bar{q}q\rangle^3$ and $\langle\frac{\alpha_sGG}{\pi}\rangle\langle\bar{q}q\rangle^2\langle\bar{q}g_s\sigma Gq\rangle$ are neglectful due to the small values.
The condensates $\langle g_s^3 GGG\rangle$, $\langle g_s^3 GGG\rangle\langle\bar{q}q\rangle$ and
 $\langle \frac{\alpha_s GG}{\pi}\rangle\langle \bar{q} g_s \sigma Gq\rangle$ are not associated with the $\frac{1}{T^2}$, and play a tiny role in determining the Borel window, and they are neglected. Furthermore, we neglect the $\langle g_s^3 GGG\rangle\langle\bar{q}q\rangle^2$ due to the small value, it is also beyond the order $\mathcal{O}(\alpha_s^1)$.

 In summary, we take into account the vacuum condensates $\langle\bar{q}q\rangle$, $\langle\frac{\alpha_sGG}{\pi}\rangle$, $\langle\bar{q}g_s\sigma Gq\rangle$, $\langle\bar{q}q\rangle^2$,  $\langle\bar{q}q\rangle\langle\frac{\alpha_sGG}{\pi}\rangle$,
  $\langle\bar{q} q\rangle\langle\bar{q}g_s\sigma Gq\rangle$, $\langle\bar{q} q\rangle^3$, $\langle\bar{q}q\rangle^2\langle\frac{\alpha_sGG}{\pi}\rangle$,
   $\langle\bar{q}g_s\sigma Gq\rangle^2$,  $\langle\bar{q} q\rangle^2\langle\bar{q}g_s\sigma Gq\rangle$,  $\langle\bar{q} q\rangle\langle\bar{q}g_s\sigma Gq\rangle^2$,
 $\langle\bar{q}q\rangle^3\langle\frac{\alpha_sGG}{\pi}\rangle$,  $\langle\bar{q}g_s\sigma Gq\rangle^3$.

Now  we  take the quark-hadron duality below the continuum thresholds  $s_0$ and perform the Borel transformation  in regard to $P^2=-p^2$ to obtain  the   QCD sum rules:
\begin{eqnarray}\label{pole+Two-QCDSR-S}
\lambda^{2}_{S}\exp\left( -\frac{M_S^2}{T^2}\right)+\int_{\Delta^2}^{s_0}ds \rho_{H,S}\exp\left( -\frac{s}{T^2}\right)
&=& \int_{9m_c^2}^{s_0}ds \rho_S(s)\exp\left( -\frac{s}{T^2}\right)\, ,
\end{eqnarray}
\begin{eqnarray}\label{pole+Two-QCDSR-A}
\lambda^{2}_{A}\exp\left( -\frac{M_A^2}{T^2}\right)+\int_{\Delta^2}^{s_0}ds \rho_{H,A}
\exp\left( -\frac{s}{T^2}\right)&=& \int_{9m_c^2}^{s_0}ds \rho_A(s)\exp\left( -\frac{s}{T^2}\right)\, ,
\end{eqnarray}
the very very lengthy  expressions of the QCD spectral densities $\rho_S(s)$ and $\rho_A(s)$ are neglected for simplicity.

We differentiate    Eqs.\eqref{pole+Two-QCDSR-S}-\eqref{pole+Two-QCDSR-A} with respect to  $\tau=\frac{1}{T^2}$, then eliminate the
 pole residues $\lambda_{S}$ and $\lambda_{A}$ and obtain the QCD sum rules for
 the masses of the triply-charmed dibaryon  states,
 \begin{eqnarray}\label{pole+Two-QCDSR-S-Deri}
 M^2_{S} &=& \frac{-\frac{d}{d\tau}\left[\int_{9m_c^2}^{s_0}ds \rho_{S}(s)\exp\left( -s\tau\right)-\int_{\Delta^2}^{s_0}ds \rho_{H,S}(s)\exp\left( -s\tau\right)\right]}{\int_{9m_c^2}^{s_0}ds \rho_{S}(s)\exp\left( -s\tau\right)-\int_{\Delta^2}^{s_0}ds \rho_{H,S}(s)\exp\left( -s\tau\right)}\, ,
 \end{eqnarray}
\begin{eqnarray}\label{pole+Two-QCDSR-A-Deri}
  M^2_{A} &=& \frac{-\frac{d}{d\tau}\left[\int_{9m_c^2}^{s_0}ds \rho_{A}(s)\exp\left( -s\tau\right)-\int_{\Delta^2}^{s_0}ds \rho_{H,A}(s)\exp\left( -s\tau\right)\right]}{\int_{9m_c^2}^{s_0}ds \rho_{A}(s)\exp\left( -s\tau\right)-\int_{\Delta^2}^{s_0}ds \rho_{H,A}(s)\exp\left( -s\tau\right)}\, .
\end{eqnarray}

If the QCD sum rules can be saturated with the scalar and axialvector dibaryon states alone,  we set $\rho_{H,S}=\rho_{H,A}=0$  in Eqs.\eqref{pole+Two-QCDSR-S}-\eqref{pole+Two-QCDSR-A-Deri}, we obtain the two sets of QCD sum rules for the dibaryon states, and refer those QCD sum rules as the QCDSR I, and refer the QCD sum rules in  Eqs.\eqref{pole+Two-QCDSR-S}-\eqref{pole+Two-QCDSR-A-Deri} as the QCDSR II.
 On the other hand, if the QCD sum rules can be saturated with the scalar and axialvector two-baryon scattering  states alone, we set $\lambda_S=\lambda_A=0$, we obtain the two QCD sum rules,
\begin{eqnarray}\label{QCDSR-kappa-S}
\int_{\Delta^2}^{s_0}ds \rho_{H,S}\exp\left( -\frac{s}{T^2}\right)&=&\kappa_S \int_{9m_c^2}^{s_0}ds \rho_S(s)\exp\left( -\frac{s}{T^2}\right)\, ,
\end{eqnarray}
\begin{eqnarray}\label{QCDSR-kappa-A}
\int_{\Delta^2}^{s_0}ds \rho_{H,A}
\exp\left( -\frac{s}{T^2}\right)&=& \kappa_A\int_{9m_c^2}^{s_0}ds \rho_A(s)\exp\left( -\frac{s}{T^2}\right)\, ,
\end{eqnarray}
where we introduce the parameters $\kappa_S$ and $\kappa_A$ to parameterize the deviations from the value 1. Thereafter, we refer the QCD sum rules in Eqs.\eqref{QCDSR-kappa-S}-\eqref{QCDSR-kappa-A} as the QCDSR III.

\section{Numerical results and discussions}

We choose  the standard values of the vacuum condensates $\langle
\bar{q}q \rangle=-(0.24\pm 0.01\, \rm{GeV})^3$,   $\langle
\bar{q}g_s\sigma G q \rangle=m_0^2\langle \bar{q}q \rangle$,
$m_0^2=(0.8 \pm 0.1)\,\rm{GeV}^2$,  $\langle \frac{\alpha_s
GG}{\pi}\rangle=(0.33\,\rm{GeV})^4 $    at the energy scale  $\mu=1\, \rm{GeV}$
\cite{Reinders85,SVZ79,Colangelo-Review}, and choose the $\overline{MS}$ mass  $m_{c}(m_c)=(1.275\pm0.025)\,\rm{GeV}$
 from the Particle Data Group \cite{PDG}, and set $m_u=m_d=0$.
 We take into account
the energy-scale dependence of  the input parameters,
\begin{eqnarray}
\langle\bar{q}q \rangle(\mu)&=&\langle\bar{q}q \rangle({\rm 1GeV})\left[\frac{\alpha_{s}({\rm 1GeV})}{\alpha_{s}(\mu)}\right]^{\frac{12}{25}}\, ,\nonumber\\
\langle\bar{q}g_s \sigma Gq \rangle(\mu)&=&\langle\bar{q}g_s \sigma Gq \rangle({\rm 1GeV})\left[\frac{\alpha_{s}({\rm 1GeV})}{\alpha_{s}(\mu)}\right]^{\frac{2}{25}}\, , \nonumber\\
m_c(\mu)&=&m_c(m_c)\left[\frac{\alpha_{s}(\mu)}{\alpha_{s}(m_c)}\right]^{\frac{12}{25}} \, ,\nonumber\\
\alpha_s(\mu)&=&\frac{1}{b_0t}\left[1-\frac{b_1}{b_0^2}\frac{\log t}{t} +\frac{b_1^2(\log^2{t}-\log{t}-1)+b_0b_2}{b_0^4t^2}\right]\, ,
\end{eqnarray}
  where $t=\log \frac{\mu^2}{\Lambda^2}$, $b_0=\frac{33-2n_f}{12\pi}$, $b_1=\frac{153-19n_f}{24\pi^2}$, $b_2=\frac{2857-\frac{5033}{9}n_f+\frac{325}{27}n_f^2}{128\pi^3}$,  $\Lambda=210\,\rm{MeV}$, $292\,\rm{MeV}$  and  $332\,\rm{MeV}$ for the flavors  $n_f=5$, $4$ and $3$, respectively  \cite{PDG,Narison-mix,Narison-Book}, and evolve all the input parameters to the best energy scales   to extract the dibaryon  masses with the flavor $n_f=4$.

In the article, we study the baryon-baryon type  six-quark states (dibaryon states) or hexaquark  states, which have three charmed quarks.
Those six-quark systems  are characterized by the effective charmed quark mass  or constituent quark mass ${\mathbb{M}}_c$
and the virtuality  $V=\sqrt{M^2_{D}-(3{\mathbb{M}}_c)^2}$, where the   $D$ denotes  the dibaryon states. We set the energy  scales of the QCD spectral densities to be $\mu=V$, it is a straight forward extension of  the energy scale formula $\mu=\sqrt{M^2_{X/Y/Z}-(2{\mathbb{M}}_c)^2}$ suggested  for the hidden-charm tetraquark molecular states to the triply-charmed dibaryon states \cite{WangZG-4-quark-mole}.
 In this article, we choose the updated value ${\mathbb{M}}_c=1.85\,\rm{GeV}$  \cite{WangZG-CPC-Y4390}, and take
 the energy scale formula,
\begin{eqnarray}\label{formula}
\mu&=&\sqrt{M^2_{D}-(3{\mathbb{M}}_c)^2}\, ,
\end{eqnarray}
 as a powerful constraint to satisfy.

The relevant baryon masses and pole residues are $m_{\Sigma_c^{++}}=2453.97\,\rm{MeV}$, $m_{\Sigma_c^{+}}=2452.9\,\rm{MeV}$, $m_{\Sigma_c^{0}}=2453.75\,\rm{MeV}$,  $m_{\Xi_{cc}^{++}}=3621.2\,\rm{MeV}$ from the Particle Data Group \cite{PDG}, $m_{\Sigma}=2.40\,\rm{GeV}$, $m_{\Xi}=3.63\,\rm{GeV}$, $\lambda_\Sigma=0.045\,\rm{GeV}^3$ and $\lambda_\Xi=0.102\,\rm{GeV}^3$ from the QCD sum rules \cite{WangZG-Sigma-PLB,WangZG-Xicc-EPJC}. In this article, we choose the values $m_{\Sigma}=2.45\,\rm{GeV}$ and $m_{\Xi}=3.62\,\rm{GeV}$.  We vary the baryon masses $m_{\Sigma}$ and $m_{\Xi}$ slightly, which leads  to neglectful  changes. In this article, we take the continuum threshold parameter  as  $\sqrt{s_0}=m_{\Sigma}+m_{\Xi}+(0.5\sim0.7)\,\rm{GeV}=6.1\pm0.6\,\rm{GeV}$ to take into account the contributions from the $\Sigma_c$ and $\Xi_{cc}$ baryon states sufficiently.

 We define the pole contributions $\rm{PC}$ as
\begin{eqnarray}
{\rm PC}&=& \frac{ \int_{9m_c^2}^{s_0} ds\,\rho_{S/A}(s)\,\exp\left(-\frac{s}{T^2}\right)}{\int_{9m_c^2}^{\infty} ds \,\rho_{S/A}(s)\,\exp\left(-\frac{s}{T^2}\right)}\, .
\end{eqnarray}
For the multiquark states, it is difficult to satisfy the pole dominance criterion. The energy scale formula, see Eq.\eqref{formula}, can enhance the pole contributions significantly, and also can improve the convergent behaviors of the operator product expansion  substantially.
If the operator product expansion is convergent, the  contributions of the higher dimensional vacuum condensates $D(n)$ with $n\geq 10$ play a minor important role,
   \begin{eqnarray}
D(n)&=& \frac{  \int_{9m_c^2}^{s_0} ds\,\rho_{S/A;n}(s)\,\exp\left(-\frac{s}{T^2}\right)}{\int_{9m_c^2}^{s_0} ds \,\rho_{S/A}(s)\,\exp\left(-\frac{s}{T^2}\right)}\, ,
\end{eqnarray}
 where the subscript $n$ in the QCD spectral densities  $\rho_{S/A;n}(s)$  denotes  the vacuum condensates  of dimension $n$.

We obtain the Borel windows and the best energy scales of the QCD spectral densities for the QCDSR I, which are  shown in Table 1, via trial  and error.  Now it is straight forward to get  the pole contributions, the pole contributions are as large as $(38-63)\%$, it is large enough to extract the dibaryon masses. In the QCD sum rules for the multiquark states, the pole contributions are usually small due to the QCD spectral densities  $\rho_{QCD}(s)\propto s^n$ with the largest value $n\geq 4$ in the zero quark mass limit \cite{WangZG-NPA-Y2175}.

\begin{figure}
\centering
\includegraphics[totalheight=7cm,width=9cm]{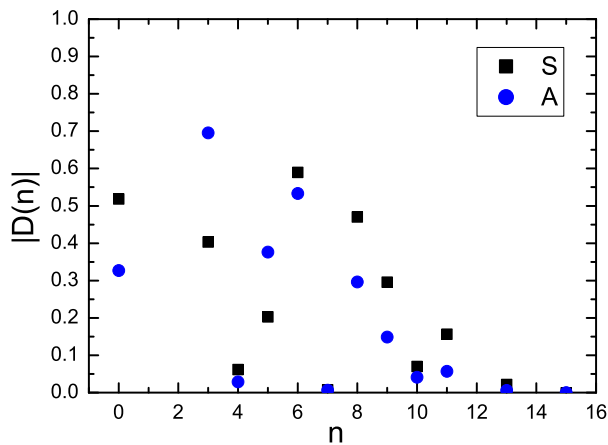}
  \caption{ The absolute values of the contributions of the vacuum condensates of dimension $n$ for  central values of the input parameters  in the QCDSR I/II, where the $S$ and $A$ represent the scalar and axialvector dibaryon states respectively.  }\label{OPE}
\end{figure}

In Fig.\ref{OPE}, we plot the absolute values of the $D(n)$ for the central values of the input parameters. Although the perturbative contributions are not large enough to dominate  the QCD sum rules,  the vacuum condensate $\langle\bar{q}q\rangle^2$ with the dimension 6 serves   as a milestone,   the absolute values of the
 contributions of the vacuum condensates with $n\geq 6$ decrease  monotonically  and quickly with  increase of the dimension except for the vacuum condensate $\langle\bar{q}q\rangle\langle \frac{\alpha_sGG}{\pi}\rangle$, which plays a tiny role. The convergent behavior of the operator product expansion is very good.

\begin{figure}
\centering
\includegraphics[totalheight=6cm,width=7cm]{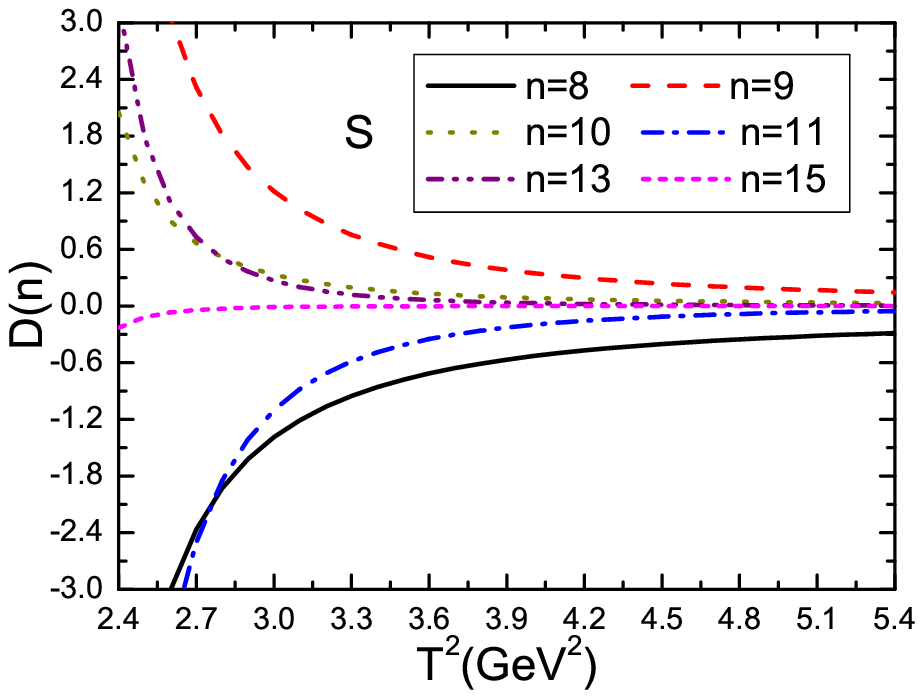}
\includegraphics[totalheight=6cm,width=7cm]{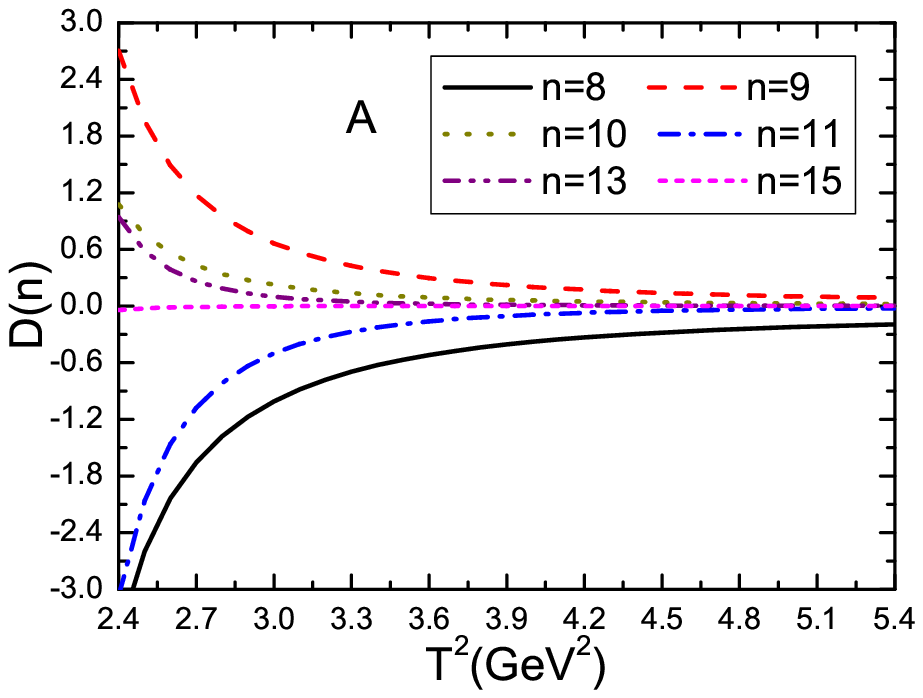}
  \caption{ The contributions of the higher dimensional vacuum condensates  with variations of the  Borel parameters $T^2$ in the QCDSR I/II, where the $S$ and $A$ represent the scalar and axialvector dibaryon states respectively.   }\label{OPE-SA}
\end{figure}

 Although the higher dimensional vacuum condensates play a minor important role in the Borel windows, they play an important role in determining  the Borel windows, we should take them into account in a consistent way. In Fig.\ref{OPE-SA}, we plot  the  values of the $D(n)$ for $n\geq8$ with variations of the Borel parameters for the central values of the other input parameters.
From Fig.\ref{OPE-SA}, we observe  that the contributions of the vacuum condensates of dimensions $8$, $9$, $10$, $11$ and $13$ are large at the region $T^2<3.9\,\rm{GeV}^2$, we should choose $T^2\geq 3.9\,\rm{GeV}^2$, while the contributions of the vacuum condensates of dimension 15  are very small in the whole regions. In fact, the vacuum condensate $\langle\bar{q}g_s\sigma Gq\rangle^3$ is the vacuum expectation of the operator of the order $\mathcal{O}(\alpha_s^{\frac{3}{2}})$, which is beyond the truncation $\mathcal{O}(\alpha_s^{k})$ with $k\leq 1$.

Now we estimate the contributions of the vacuum condensates $\langle\bar{q}q\rangle^3\langle g_s^3 GGG\rangle$, which are also of dimension $15$.
If each light quark propagator contributes a quark-antiquark pair, which  forms the quark condensate, $S_{ij}(x)\to -\frac{1}{12}\langle\bar{q}q\rangle\delta_{ij}$, we can write down  the corresponding contribution $\Pi_S(p)$ (the corresponding contribution $\Pi_{\mu\nu}(p)$ can be treated  similarly),

\begin{eqnarray}\label{PI-S-qqqqqq}
\Pi_S(p)&=&-\frac{2}{27}\langle\bar{q}q\rangle^3\,\Pi_{ccc}(p)+\cdots \, ,
\end{eqnarray}
where
\begin{eqnarray}
\Pi_{ccc}(p)&=&i\,\varepsilon^{lmn}\varepsilon^{l^{\prime}m^{\prime}n}\int d^4x\, e^{ip\cdot x} {\rm Tr}\left[\gamma_\beta  S_c^{mm^\prime}(x)   \gamma^{\beta} C S_c^{ll^\prime T}(x)C\right]
 {\rm Tr} \left[ C S_c^{kk T}(x)C \right]     \, .
\end{eqnarray}
The correlation function  $\Pi_{ccc}(p)$ has three heavy  quark propagators, just like in the QCD sum rules for the triply-heavy baryon states \cite{ZhangJR-triply}.
The vacuum condensates in the correlation functions $\Pi_S(p)$ and $\Pi_{ccc}(p)$ have the following relations,
\begin{eqnarray}
\langle\bar{q}q\rangle^3 &\leftrightarrow & {\rm perturbative\,\,\,term}\, , \nonumber\\
\langle\bar{q}q\rangle^3 \langle\frac{\alpha_sGG}{\pi}\rangle&\leftrightarrow & \langle\frac{\alpha_sGG}{\pi}\rangle\, , \nonumber\\
\langle\bar{q}q\rangle^3 \langle g_s^3GGG\rangle&\leftrightarrow & \langle g_s^3GGG\rangle\, .
\end{eqnarray}
In the QCD sum rules for the triply-heavy baryon states, we usually take into account the perturbative terms and the gluon condensates $\langle\frac{\alpha_sGG}{\pi}\rangle$, and neglect  the contributions of the three-gluon condensates due to their small values \cite{ZhangJR-triply}.
We expect that the  contributions of the three-gluon condensates in the $\Pi_{ccc}(p)$ can be  neglected, just like in the case of the triply-heavy baryon states, therefore, the corresponding vacuum condensates  $\langle\bar{q}q\rangle^3\langle g_s^3 GGG\rangle$ in the $\Pi_S(p)$ can also  be neglected.

If we want to calculate the three-gluon condensates, we should take into account the three-gluon operators neglected in the full propagators in Eqs.\eqref{LQuarkProg}-\eqref{HQuarkProg}.
In Ref.\cite{WangZG-tensor-M}, we calculate the contributions of the vacuum condensates up to dimension-6
in the operator product expansion, study the masses and decay constants of the heavy tensor
mesons with the QCD sum rules. In calculations, we observe that the  contributions come from the gluon condensates and three-gluon condensates
are about $10\%$ and $0.2\%$ respectively for the charmed mesons. It is indeed that the contributions of the three-gluon condensates are much smaller than that of the gluon condensates.

 From Fig.\ref{OPE}, we can see that the contributions of the vacuum condensates $\langle\bar{q} q\rangle\langle\bar{q}g_s\sigma Gq\rangle^2$ and
 $\langle\bar{q}q\rangle^3\langle\frac{\alpha_sGG}{\pi}\rangle$, which are  of dimension $13$ and have two gluon operators,  are tiny. We can draw the conclusion tentatively that the contributions of the vacuum condensates $\langle\bar{q}g_s\sigma Gq\rangle^3$, $\langle\frac{\alpha_sGG}{\pi}\rangle\langle\bar{q}q\rangle^2\langle\bar{q}g_s\sigma Gq\rangle$ and  $\langle\bar{q}q\rangle^3\langle g_s^3 GGG\rangle$, which   are of dimension $15$ and have three gluon operators, are  even tiny. It is indeed that the contributions of the vacuum condensates $\langle\bar{q}g_s\sigma Gq\rangle^3$, which are of dimension $15$, are even tiny, see Figs.\ref{OPE}-\ref{OPE-SA}. The higher dimensional vacuum condensates play a minor important role in the Borel windows, the operator product expansion is well convergent.

\begin{figure}
\centering
\includegraphics[totalheight=6cm,width=7cm]{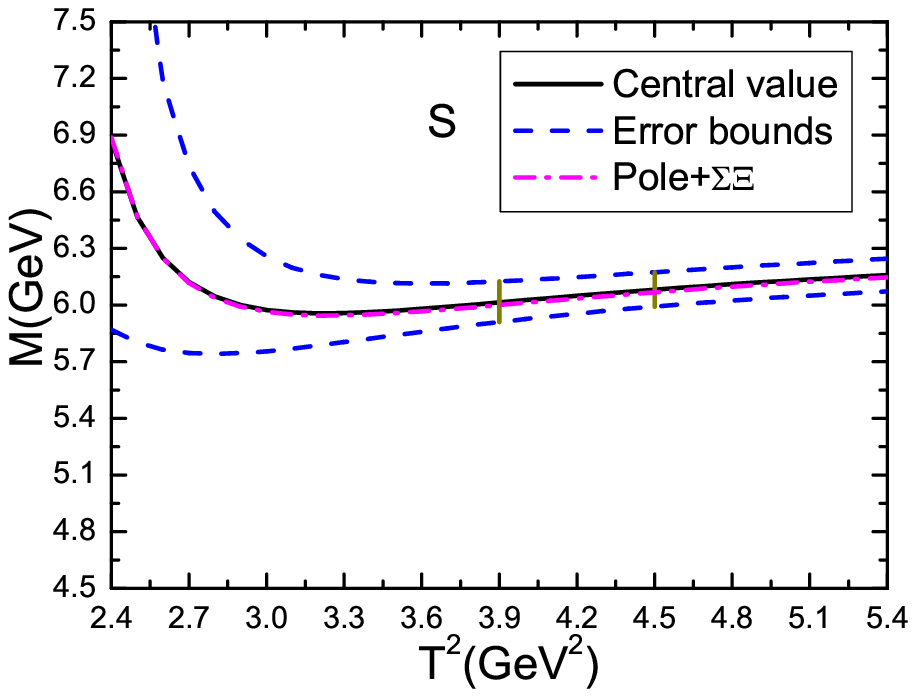}
\includegraphics[totalheight=6cm,width=7cm]{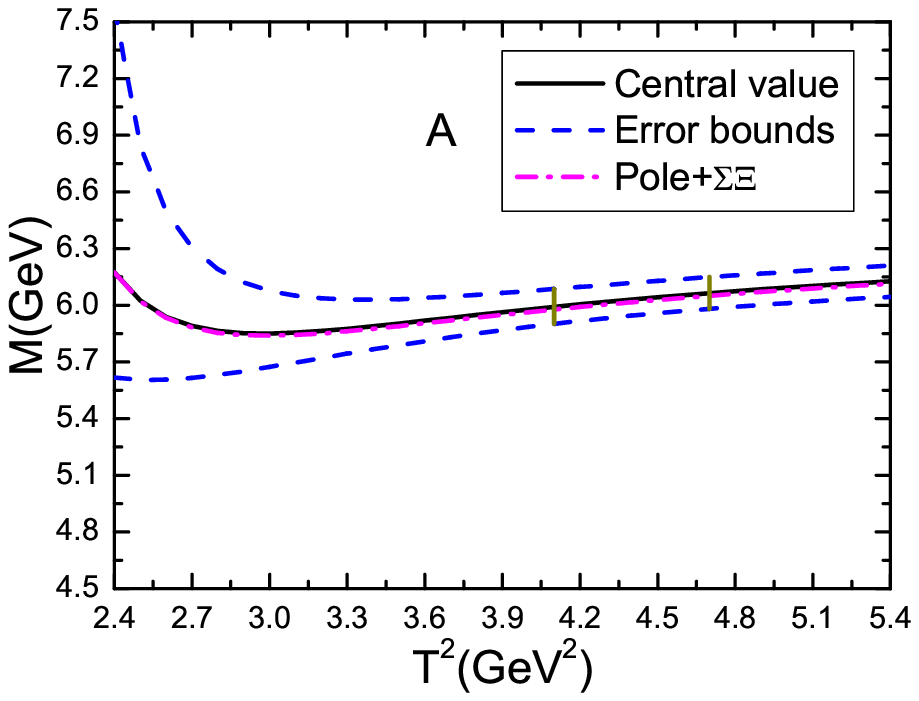}
  \caption{ The masses of the dibaryon states with variations of the  Borel parameters $T^2$ from the QCDSR I, where the $S$ and $A$ represent the scalar and axialvector dibaryon states respectively,  the Pole+$\Sigma\Xi$ corresponds to the results from the QCDSR II.   }\label{mass-SA}
\end{figure}

\begin{figure}
\centering
\includegraphics[totalheight=6cm,width=7cm]{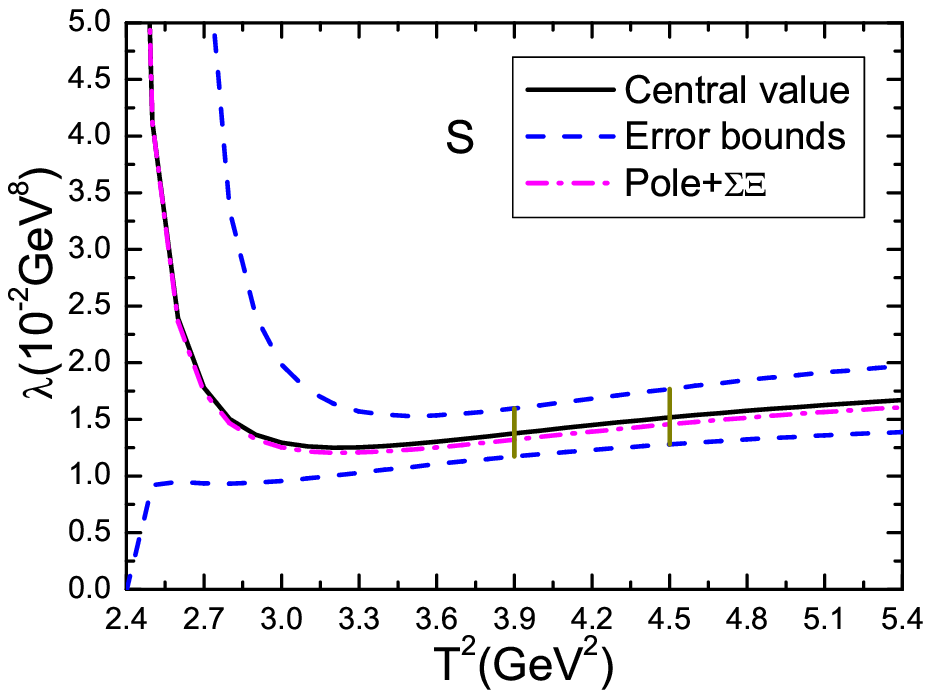}
\includegraphics[totalheight=6cm,width=7cm]{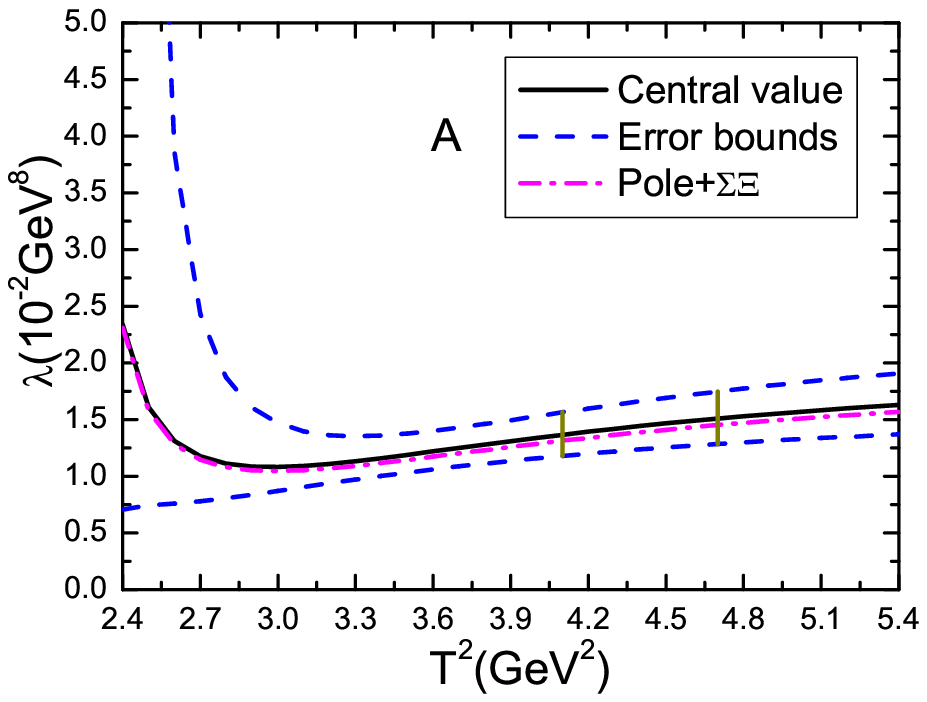}
  \caption{ The pole residues of the dibaryon states with variations of the  Borel parameters $T^2$ from the QCDSR I, where the $S$ and $A$ represent the scalar and axialvector dibaryon states respectively, the Pole+$\Sigma\Xi$ corresponds to the results from the QCDSR II.  }\label{residue-SA}
\end{figure}

Now we take  into account all uncertainties of the input parameters,
and obtain the values of the masses and pole residues of
 the   triply-charmed dibaryon states from the QCDSR I and II respectively, which are  shown explicitly in Table \ref{BCEPMR} and Figs.\ref{mass-SA}-\ref{residue-SA}, where the regions between the two vertical lines are the Borel windows.
 In Figs.\ref{mass-SA}-\ref{residue-SA}, we plot the masses and pole residues of the triply-charmed dibaryon states in  much larger  ranges than the Borel windows.
 Flat platforms  appear at the Borel windows both for the masses and pole residues, it is reliable to extract the dibaryon masses. From Table \ref{BCEPMR}, we can see that the central values of the dibaryon masses from the QCDSR I satisfy the energy scale formula  $\mu=\sqrt{M^2_{D}-(3{\mathbb{M}}_c)^2}$. If we take into account the
 two-baryon  scattering states, we obtain a slightly smaller masses and pole residues for the dibaryon states from the QCDSR II, see Table \ref{BCEPMR} and Figs.\ref{mass-SA}-\ref{residue-SA}. The effects of the two baryon scattering states are rather small and can be neglected safely.

In Ref.\cite{Latt-dibaryon-PRL}, the lattice QCD calculations indicate that the color-singlet-color-singlet type currents with $J^P=1^+$ have non-vanishing couplings with the heavy dibaryon states, which lie below the corresponding two-baryon thresholds. In Ref.\cite{ccc-dibaryon}, the calculations based on the one-boson-exchange model indicate that there exist attractive  interactions between the $\Xi_{cc}$ and $\Sigma_c$ in the channels $I(J^P)=\frac{1}{2}(0^+)$, $\frac{1}{2}(1^+)$ and $\frac{3}{2}(0^+)$.
In Ref.\cite{PanLDXX}, the heavy antiquark-diquark symmetry is used to study the mass-splitting between the $J^P=0^+$ and $1^+$ $\Xi_{cc}\Sigma_c$ dibaryon states in a model-independent way. In the present work, the predictions of the central values $M_S=6.05\,\rm{GeV}$ ($6.04\,\rm{GeV}$) and $M_A=6.03\,\rm{GeV}$ ($6.02\,\rm{GeV}$) from  the QCDSR I (QCDSR II) are consistent with the previous works \cite{PanLDXX,ccc-dibaryon,Latt-dibaryon-PRL}.

\begin{table}
\begin{center}
\begin{tabular}{|c|c|c|c|c|c|c|c|}\hline\hline
$J^P$   &$T^2(\rm{GeV}^2)$   &$\sqrt{s_0}(\rm{GeV})$   &$\mu(\rm{GeV})$  &pole          &$M(\rm{GeV})$  &$\lambda(\rm{GeV}^8)$ \\ \hline

$0^+$(I)   &$3.9-4.5$           &$6.7\pm0.1$              &$2.4$            &$(38-63)\%$   &$6.05\pm0.13$  &$(1.45\pm0.30)\times10^{-2}$   \\
$0^+$(II)  &                    &                         &                 &              &$6.04\pm0.13$  &$(1.39\pm0.30)\times10^{-2}$   \\ \hline

$1^+$(I)   &$4.1-4.7$           &$6.7\pm0.1$              &$2.4$            &$(38-62)\%$   &$6.03\pm0.13$  &$(1.44\pm0.29)\times10^{-2}$   \\
$1^+$(II)  &                    &                         &                 &              &$6.02\pm0.13$  &$(1.39\pm0.28)\times10^{-2}$   \\ \hline\hline
\end{tabular}
\end{center}
\caption{ The Borel parameters, continuum threshold parameters,   energy scales, pole contributions,   masses and pole residues for the
 triply-charmed   dibaryon  states from the QCD sum rules  I and II. }\label{BCEPMR}
\end{table}

\begin{figure}
\centering
\includegraphics[totalheight=7cm,width=9cm]{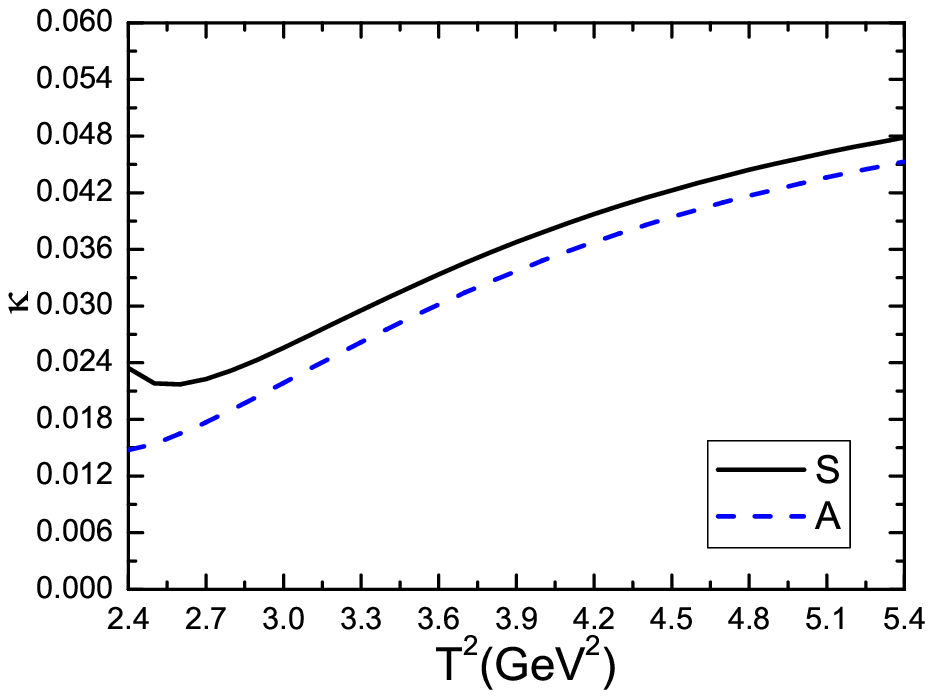}
  \caption{  The parameters $\kappa$ with variations of the  Borel parameters $T^2$ from the QCDSR III, where the $S$ and $A$ represent the scalar and axialvector dibaryon states respectively. }\label{kappa-SA}
\end{figure}

In Fig.\ref{kappa-SA}, we plot the parameters $\kappa_S$ and $\kappa_A$ with variations of the Borel parameters for the central values of the input
parameters shown in Table \ref{BCEPMR}. From the figure, we can see that the values $\kappa_S\ll1$ and $\kappa_A\ll1$, furthermore, the values of the $\kappa_S$ and $\kappa_A$ increase monotonically and quickly with increase of the Borel parameters, no flat platforms  can be obtained. If we choose smaller energy scale, say $\mu=1\,\rm{GeV}$ (which is chosen in the QCD sum rules for the $\Sigma_c$ and $\Xi_{cc}$ \cite{WangZG-Sigma-PLB,WangZG-Xicc-EPJC}), we can obtain very large values for the $\kappa_S$ and $\kappa_A$ with poor pole contributions, the values of the $\kappa_S$ and $\kappa_A$ decrease monotonically  and quickly with  increase of the Borel parameters, on the other hand, the convergent behaviors of the operator product expansion are very bad.
We can draw the conclusion tentatively that the two-baryon scattering states cannot saturate the QCD sum rules and play a minor important role and can be neglected, which approves the observation obtained in Sect.2,  the molecular states begin to receive contributions at the order $\mathcal{O}(\alpha_s^0)$, not at the order $\mathcal{O}(\alpha_s^2)$.

\section{Conclusion}
In this article,  we construct the color-singlet-color-singlet type currents to interpolate the scalar and axialvector $\Xi_{cc}\Sigma_c$ dibaryon states, and study them with QCD sum rules in details by carrying out the operator product expansion up to the vacuum condensates of dimension 15. At the hadron side, we take into account both the contributions of the  dibaryon states and two-baryon scattering sates,  and explore the  existence or nonexistence of the $\Xi_{cc}\Sigma_c$ dibaryon states in  three cases. As there exist three valance $c$-quarks, the QCD sum rules are sensitive to the  $c$-quark mass or the energy scale of the QCD spectral densities, we determine the best energy scales with the  energy scale formula $\mu=\sqrt{M^2_{D}-(3{\mathbb{M}}_c)^2}$. The numerical results indicate that the two-baryon scattering states cannot saturate the QCD sum rules and play a minor important role, the dominant contributions come from the dibaryon states. Or it is necessary to introduce the dibaryon states to embody the net effects.  The color-singlet-color-singlet type currents couple potentially to the molecular states, not to the two-particle scattering states. In the operator product expansion, the molecular states begin to receive contributions at the order $\mathcal{O}(\alpha_s^0)$, not at the order $\mathcal{O}(\alpha_s^2)$.
We  can search for the triply-charmed  dibaryon states  at the LHCb, Belle II,  CEPC, FCC, ILC  in the future.

\section*{Acknowledgements}
This  work is supported by National Natural Science Foundation, Grant Number  11775079.


\begin{thebibliography}{99}


\bibitem{d2380-establish} P. Adlarson et al, Phys. Rev. {\bf C90} (2014)  035204.

\bibitem{d2380-pipi} P. Adlarson  et al, Phys. Rev. Lett. {\bf 106} (2011) 242302;
P. Adlarson  et al, Phys. Rev. Lett. {\bf 112} (2014)  202301.


\bibitem{dibaryon-review} H. Clement, Prog. Part. Nucl. Phys. {\bf 93} (2017) 195.


\bibitem{X3872-2003} S. K. Choi   et al, Phys. Rev. Lett. {\bf 91} (2003) 262001.


\bibitem{GFK-RMP} F. K. Guo, C. Hanhart, U. G. Meissner, Q. Wang, Q. Zhao and B. S. Zou, Rev. Mod. Phys. {\bf 90} (2018)  015004.



\bibitem{LHCb-4380} R. Aaij  et al, Phys. Rev. Lett. {\bf 115} (2015) 072001.

\bibitem{LHCb-Pc4312} R. Aaij et al, Phys. Rev. Lett. {\bf 122} (2019) 222001.

\bibitem{PanLDXX} Y. W. Pan, M. Z. Liu, F. Z. Peng, M. S. Sanchez, L. S. Geng and M. P. Valderrama, arXiv:1907.11220.

\bibitem{ccc-dibaryon} R. Chen, F. L. Wang, A. Hosaka and X. Liu, Phys. Rev. {\bf D97} (2018)  114011.


\bibitem{Heavy-dibaryon-hadron} W. Meguro, Y. R. Liu and M. Oka, Phys. Lett. {\bf B704} (2011) 547;
L. Meng, N. Li and S. L. Zhu, Phys. Rev. {\bf D95} (2017)  114019.


\bibitem{Latt-dibaryon-PRL} P. Junnarkar and N. Mathur, Phys. Rev. Lett. {\bf 123} (2019)  162003.


\bibitem{Heavy-dibaryon-quark} H. Huang, J. Ping and F. Wang, Phys. Rev. {\bf C89} (2014)  035201;
J. Vijande, A. Valcarce, J. M. Richard and P. Sorba, Phys. Rev. {\bf D94} (2016)  034038.

\bibitem{LHCb-Xicc} R. Aaij  et al,  Phys. Rev. Lett. {\bf 119} (2017)  112001.



\bibitem{Oka-H-dibaryon} N. Kodama, M. Oka and T. Hatsuda, Nucl. Phys. {\bf A580} (1994) 445.

\bibitem{D2380-CHHX} H. X. Chen, E. L. Cui, W. Chen, T. G. Steele  and S. L. Zhu, Phys. Rev. {\bf C91} (2015)  025204.


\bibitem{QCDSR-4-quark-mass} R. D. Matheus, S. Narison, M. Nielsen and J. M. Richard, Phys. Rev. {\bf D75} (2007) 014005;
Z. G. Wang, Eur. Phys. J. {\bf C63} (2009) 115;
W. Chen and S. L. Zhu, Phys. Rev. {\bf D83} (2011) 034010;
J. R. Zhang, M. Zhong and M. Q. Huang, Phys. Lett. {\bf B704} (2011) 312;
J. R. Zhang,  Phys. Rev. {\bf D87} (2013)  116004;
C. F. Qiao and L. Tang, Eur. Phys. J. {\bf C74} (2014) 2810;
Z. G. Wang, Commun. Theor. Phys. {\bf 63} (2015) 466;
Z. G. Wang and Y. F Tian, Int. J. Mod. Phys. {\bf A30} (2015) 1550004;
Z. G. Wang,  Eur. Phys. J. {\bf C79} (2019)  29.

\bibitem{WangHuangtao-PRD} Z. G. Wang and T. Huang, Phys. Rev. {\bf D89} (2014)  054019.


\bibitem{QCDSR-4-quark-width} F. S. Navarra and M. Nielsen, Phys. Lett. {\bf B639} (2006) 272;
J. M. Dias, F. S. Navarra, M. Nielsen, C. M. Zanetti, Phys. Rev. {\bf D88} (2013)  016004;
W. Chen, T. G. Steele, H. X. Chen and S. L. Zhu, Eur. Phys. J. {\bf C75} (2015)  358;
Z. G. Wang and T. Huang, Nucl. Phys. {\bf A930} (2014) 63;
S. S. Agaev, K. Azizi and H. Sundu, Phys. Rev. {\bf D93} (2016)  074002;
Z. G. Wang and J. X. Zhang, Eur. Phys. J. {\bf C78} (2018)  14;
Z. G. Wang, Int. J. Mod. Phys. {\bf A34} (2019) 1950110.


\bibitem{WangZG-4-quark-mole} Z. G. Wang and T. Huang, Eur. Phys. J. {\bf C74} (2014)  2891;
Z. G. Wang, Eur. Phys. J. {\bf C74} (2014)  2963.


\bibitem{WangZG-CPC-Y4390} Z. G.  Wang, Chin. Phys. {\bf C41} (2017)  083103.


\bibitem{QCDSR-5-quark-mole} H. X. Chen, W. Chen, X. Liu, T. G. Steele, S. L. Zhu, Phys. Rev. Lett. {\bf 115} (2015)  172001;
K. Azizi, Y. Sarac and H. Sundu, Phys. Rev. {\bf D95} (2017)  094016.

\bibitem{QCDSR-WangZG-5-quark-mole}  Z. G. Wang, Int. J. Mod. Phys. {\bf A34} (2019)  1950097.


\bibitem{QCDSR-5-quark-penta} Z. G. Wang, Eur. Phys. J. {\bf C76} (2016)  70;
Z. G.  Wang and T. Huang, Eur. Phys. J. {\bf C76} (2016)  43.


\bibitem{Chu-Sheng-PRD} W. Lucha, D. Melikhov and H. Sazdjian, Phys. Rev. {\bf D100} (2019)  014010;
W. Lucha, D. Melikhov and H. Sazdjian, Phys. Rev. {\bf D100} (2019)  074029.

\bibitem{Chu-Sheng-EPJC} W. Lucha, D. Melikhov and H. Sazdjian, Eur. Phys. J. {\bf C77} (2017)  866.

\bibitem{Landau} L. D. Landau, Nucl. Phys. {\bf 13} (1959) 181.


\bibitem{Reinders85} L. J. Reinders, H. Rubinstein and S. Yazaki, Phys. Rept. {\bf 127} (1985) 1.


\bibitem{Pascual-1984} P. Pascual and R. Tarrach, ``QCD: Renormalization for the practitioner", Springer Berlin Heidelberg (1984).


\bibitem{GuoFK-Hadron2017} F. K. Guo, PoS Hadron2017 (2018) 015;
F. K. Guo, X. H. Liu and S. Sakai,  Prog. Part. Nucl. Phys. {\bf 112} (2020) 103757.


\bibitem{WangZG-Landau} Z. G. Wang, Phys. Rev. {\bf D101} (2020)  074011.


\bibitem{SVZ79} M. A. Shifman, A. I. Vainshtein and V. I. Zakharov, Nucl. Phys. {\bf B147} (1979) 385; Nucl. Phys. {\bf B147} (1979) 448.


\bibitem{Colangelo-Review}  P. Colangelo and A. Khodjamirian, hep-ph/0010175.

\bibitem{PDG}   C. Patrignani et al, Chin. Phys. {\bf C40} (2016)  100001.

\bibitem{Narison-mix} S. Narison and R. Tarrach, Phys. Lett. {\bf 125 B} (1983) 217.

\bibitem{Narison-Book} S. Narison, ``QCD as a theory of hadrons from partons to confinement", Camb. Monogr. Part. Phys. Nucl. Phys. Cosmol. {\bf 17} (2007) 1.

\bibitem{WangZG-Sigma-PLB} Z. G. Wang,  Phys. Lett. {\bf B685} (2010) 59.

\bibitem{WangZG-Xicc-EPJC} Z. G. Wang, Eur. Phys. J. {\bf C78} (2018)  826.


\bibitem{WangZG-NPA-Y2175} Z. G. Wang, Nucl. Phys. {\bf A791} (2007) 106.

\bibitem{ZhangJR-triply} J. R. Zhang and M. Q. Huang, Phys. Lett. {\bf B674} (2009) 28;
Z. G. Wang, Commun. Theor. Phys. {\bf 58} (2012) 723;
T. M. Aliev, K. Azizi and M. Savci, J. Phys. {\bf G41} (2014) 065003.

\bibitem{WangZG-tensor-M} Z. G. Wang and Z. Y. Di, Eur. Phys. J. {\bf A50} (2014) 143.

\end{thebibliography}
\end{document}